\documentclass[aps,prb,a4paper,twocolumn,showpacs,showkeys,fleqn,floatfix,superscriptaddress]{revtex4-2}

\usepackage{graphicx}%for pdf-figures
\usepackage{color}
\usepackage{amsmath}
\usepackage{multirow}

\newcommand{\figwidth}{\columnwidth}

\newcommand{\vect}[1]{\mathbf{#1}}
\newcommand{\nabok}{KCu$_7$(TeO$_4$)(SO$_4$)$_5$Cl}
\newcommand{\kagome}{kagom\'{e}}
\newcommand{\Tpeak}{T_\textrm{peak}}
\newcommand{\degr}{\ensuremath{{}^\circ}}
\newcommand{\Tafe}{T_\textrm{AFE}}
\newcommand{\TN}{T_\textrm{N}}
\newcommand{\Cp}{C_\textrm{p}}
\newcommand{\cuso}{CuSO$_4 \cdot$5H$_2$O}
\newcommand{\abs}[1]{\left|#1\right|}

\begin{document}

\title{Static and resonant properties of decorated square \kagome{} lattice compound \nabok{}}

\author{M.M.~Markina}
\email{markina@lt.phys.msu.ru}
\affiliation{M.V.~Lomonosov Moscow State University, Moscow 119991, Russia}
\affiliation{National University of Science and Technology ``MISiS'', Moscow 119049, Russia}

\author{P.S.~Berdonosov}
\affiliation{M.V.~Lomonosov Moscow State University, Moscow 119991, Russia}
\affiliation{National University of Science and Technology ``MISiS'', Moscow 119049, Russia}

\author{T.M.~Vasilchikova}
\affiliation{M.V.~Lomonosov Moscow State University, Moscow 119991, Russia}
\affiliation{National University of Science and Technology ``MISiS'', Moscow 119049, Russia}

\author{K.V.~Zakharov}
\affiliation{M.V.~Lomonosov Moscow State University, Moscow 119991, Russia}

\author{A.F.~Murtazoev}
\affiliation{M.V.~Lomonosov Moscow State University, Moscow 119991, Russia}

\author{V.A.~Dolgikh}
\affiliation{M.V.~Lomonosov Moscow State University, Moscow 119991, Russia}

\author{A.V.~Moskin}
\affiliation{M.V.~Lomonosov Moscow State University, Moscow 119991, Russia}
\affiliation{National University of Science and Technology ``MISiS'', Moscow 119049, Russia}

\author{V.N.~Glazkov}
\affiliation{P.L.~Kapitza Institute for Physical Problems, RAS, Moscow  119334, Russia}

\author{A.I.~Smirnov}
\affiliation{P.L.~Kapitza Institute for Physical Problems, RAS, Moscow  119334, Russia}

\author{A.N.~Vasiliev}
\affiliation{M.V.~Lomonosov Moscow State University, Moscow 119991, Russia}
\affiliation{National University of Science and Technology ``MISiS'', Moscow 119049, Russia}

\begin{abstract}

The magnetic subsystem of nabokoite, \nabok{}, is constituted by copper ions forming a buckled square \kagome{} lattice decorated by quasi-isolated ions. This combination determines peculiar physical properties of this compound evidenced in electron spin resonance (ESR) spectroscopy, dielectric permittivity $\varepsilon$, magnetization $M$ and specific heat $\Cp$ measurements. At lowering temperature, the magnetic susceptibility $\chi = M/H$ passes through a broad hump  inherent for low-dimensional magnetic systems at about 150 K and a sharp peak at antiferromagnetic phase transition at $\TN = 3.2 $~K. The $\Cp(T,H)$ curves demonstrate  additional peak-like anomaly at $\Tpeak= 5.7$~K robust to magnetic field. The latter can be ascribed to low-lying singlet excitations filling the singlet-triplet gap in magnetic excitation spectrum of the square \kagome{} lattice [J.~Richter, O.~Derzhko and J.~Schnack, Phys. Rev. B \textbf{105}  (2022) 144427]. ESR spectroscopy provides indications that antiferromagnetic structure below $\TN$ is non-collinear. Separate issue is the observation of antiferroelectric-type behavior in $\varepsilon$ at low temperatures, which tentatively reduces the symmetry and partially lifts frustration of magnetic interactions of decorating copper ions with buckled square \kagome{} lattice. These complex thermodynamic and resonant properties signal the presence of two weakly coupled magnetic subsystems in nabokoite, namely a  spin-liquid in square \kagome{} lattice layers and an antiferromagnet represented by decorating ions. \end{abstract}

\maketitle

\section{Introduction}
A low-dimensional magnetic
materials will play key role in the future electronic and spintronic devices. Depending on magnetic ions bonds topology, the resulting properties of
material may show different magnetic behavior. Layered 2-dimensional systems attract the attention due to the possible spin frustrations and
formation of exotic spin states. In 2001, Siddharthan and Georges introduced a two-dimensional network of corner-sharing triangles with square \kagome{} lattice (SKL) geometry \cite{ref1}. In variance with \kagome{} lattice, which is a two-dimensional network of corner-sharing triangles with hexagonal voids \cite{ref2}, SKL is a two-dimensional network of corner-sharing triangles with alternative square and octagonal voids. There are two non-equivalent positions for the magnetic ions, $\alpha$ and $\beta$, being in ratio two to one. The shape of the unit cell resembles shuriken, a Ninja concealed weapon of ancient time, which explains an alternative naming of this type of lattice as a shuriken lattice \cite{ref3}. It has been shown numerically that the properties of a square \kagome{} lattice antiferromagnet should be similar to the properties of an antiferromagnet on a traditional  \kagome{} lattice \cite{ref4}. For the spin-1/2 Heisenberg case with equal exchange interactions between all sites, a spin-liquid ground state has been predicted with a triplet gap filled by a continuum of low-lying singlet states \cite{ref5}.

For a long time, the studies of the basic properties of compounds with SKL or its derivatives were carried out exclusively by theoretical methods \cite{ref8,ref9,ref10,ref11,ref12,ref13,ref14,ref15,ref16,ref17,ref18,ref19,ref20}, albeit the Earth sciences provide these patterns in some rare minerals. Among these minerals are nabokoite \nabok{} \cite{ref21}, atlasovite KCu$_6$FeBiO$_4$(SO$_4$)$_5$Cl \cite{ref22}, elasmochloite Na$_3$Cu$_6$BiO$_4$(SO$_4$)$_5$ \cite{ref23} and favreauite PbCu$_6$BiO$_4$(SeO$_3$)$_4$(OH)H$_2$O \cite{ref24}. Intrinsic properties of SKL in natural nabokoite and atlasovite are masked by the presence of magnetic ions extraneous to this network.

Recently, the iron-free sibling of atlasovite, KCu$_6$AlBiO$_4$(SO$_4$)$_5$Cl, has been synthesized and thoroughly investigated in measurements of thermodynamics, muon spin relaxation and neutron scattering \cite{ref25}. It has been established that down to 58~mK this compound persists in a gapless quantum spin liquid state. Additionally, a novel sodium bismuth oxocuprate phosphate chloride, Na$_6$Cu$_7$BiO$_4$(PO$_4$)$_4$Cl$_3$, containing both square \kagome{} layers and interlayer Cu$^{2+}$ ions has been synthesized by hydrothermal technique \cite{ref26}. This material shows no magnetic ordering down to 50~mK forming quantum spin liquid state similar to KCu$_6$AlBiO$_4$(SO$_4$)$_5$Cl \cite{ref27}. Here we present studies of static and resonant properties of synthetic nabokoite \nabok{}.

\section{Synthesis and crystal structure}

\begin{figure}[t]
  \centering
  \includegraphics[width=\figwidth]{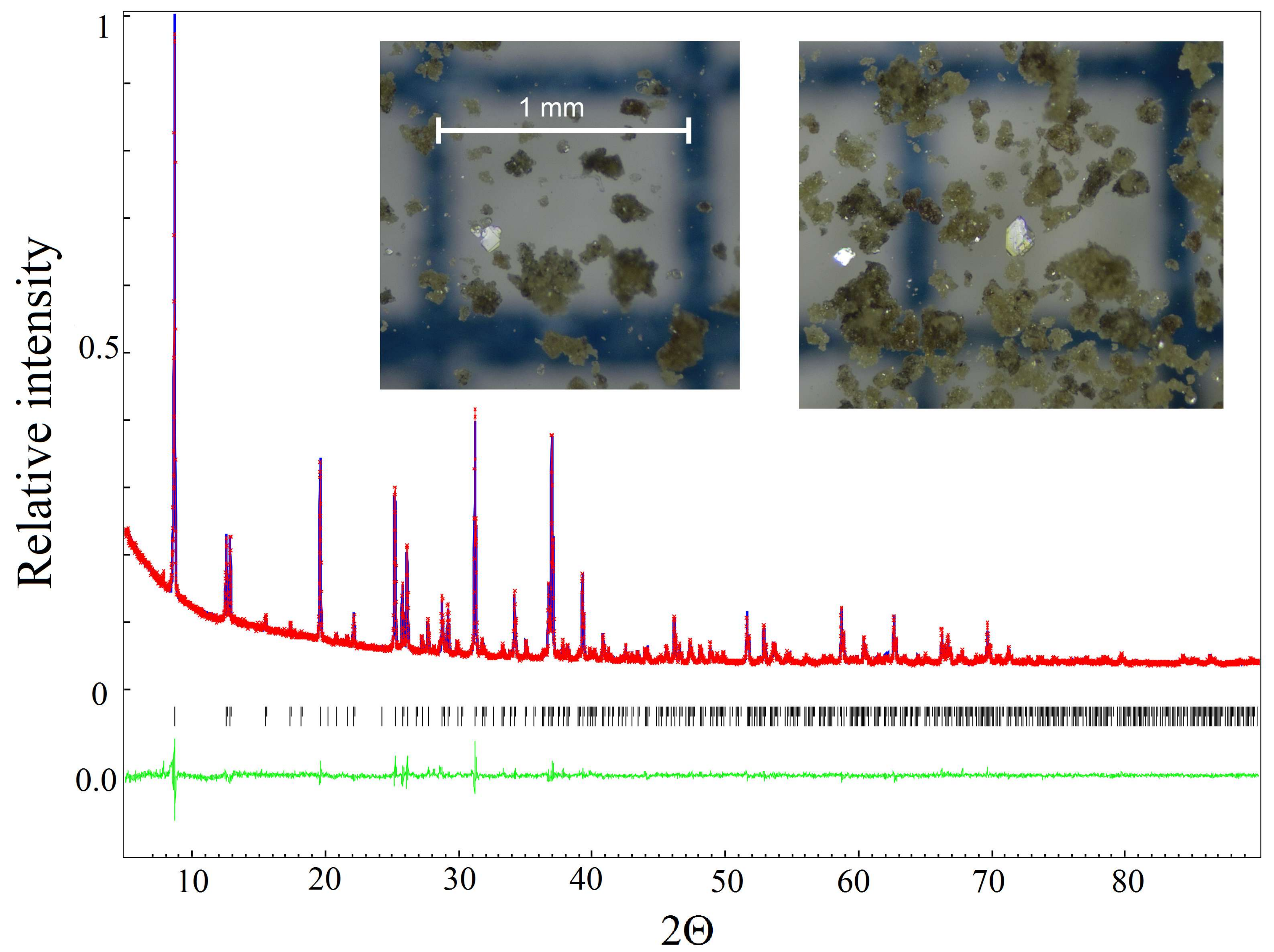}
  \caption{(color online) Rietveld plot for \nabok{} structure refinement. Red line represents experimental data, blue one --- calculated data, the difference data are depicted in green. Bragg positions are shown in black. Inset: photographs of \nabok{} sample.}\label{fig:fig1-rietveld}
\end{figure}

\begin{table}
  \centering
  \caption{Structural data for synthetic \nabok{} and mineral nabokoite}\label{tab:tab1}

\begin{tabular}{ccc}
\hline
Compound&\multicolumn{2}{c}{\nabok{}}\\
&synthetic (this work)&mineral \cite{ref21}\\
\hline
Temperature, K&293&293\\
Z&4&4\\
Symmetry&tetragonal&tetragonal\\
Space group&P4/ncc, No.130&P4/ncc, No.130\\
a, \AA{}&9.79668(7)&9.833(1)\\
c, \AA{}&20.5185(2)&20.591(2)\\
V, \AA{}$^3$&1969.26(3)&1990.90\\
$D_\textrm{calcd}$, g$\cdot$cm$^{-3}$&4.018&\\
$\lambda(\textrm{Cu K}_{\alpha 1}/\textrm{Cu K}_{\alpha 2})$, \AA{}&1.54056 / 1.54439&\\
Number of points&7008&\\
$2\Theta_\textrm{min}/(2\Theta_\textrm{max})$&5.000/89.997&\\
$Rp$&0.0309&\\
$wRp$&0.0405&\\
$wRp_\textrm{exp}$&0.0265&\\
GOF&1.53&\\
\hline
\end{tabular}

\end{table}

\begin{figure}[t!]
  \centering
  \includegraphics[width=0.7\columnwidth]{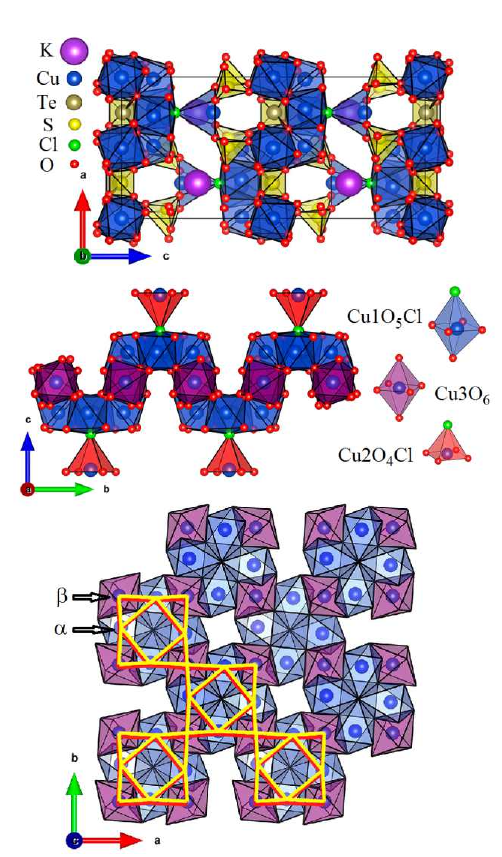}
  \caption{(color online) (a): Crystal structure of nabokoite, \nabok{}, in polyhedral representation visualized using VESTA software \cite{ref29}. K$^+$ ions are shown by the isolated spheres. (b): Buckled square \kagome{} lattice is represented by edge sharing and vertex-sharing Cu1O$_5$Cl bipyramids and Cu3O$_6$ octahedra. Dangling copper ions are situated within Cu2O$_4$Cl square pyramids. (c): Buckled layer of Cu1 and Cu3 ions viewed along the c axis. SKL is highlighted by solid lines, $\alpha$ and $\beta$ sites of SKL lattice are marked for clarity. }\label{fig:fig2-structure}
\end{figure}

Anhydrous CuSO$_4$ has been obtained by dehydration of CuSO$_4 \cdot$5H$_2$O while TeO$_2$ has been prepared by decomposition of H$_2$TeO$_4$ at 650\degr C for 50 hours. \nabok{} has been synthesized by the solid-phase ampoule method from 5~mmol CuSO$_4$, 2~mmol CuO, 1~mmol TeO$_2$ and 1~mmol KCl. All operations with anhydrous copper sulfate and homogenization have been carried in a dry box purged with argon. This mixture has been loaded into an evacuated and sealed quartz ampoule, placed in the furnace heated up to 500\degr{}C at a rate of 1.4 \degr{}C/min and kept at this temperature for a week. As a result, the sample has been sintered as small dark green crystals of irregular shape with dimensions of about a fraction of a millimeter, as shown in the inset to Fig.~\ref{fig:fig1-rietveld}. Purity of sample has been confirmed by powder XRD on BRUKER D8 Advance diffractometer (Cu K$\alpha$, $\lambda$ = 1.54056, 1.54439 \AA{}, LYNXEYE detector) using ICDD PDF2 file as a reference. The results of Rietveld refinement are shown in Fig.~\ref{fig:fig1-rietveld} and the crystal structure parameters determined using JANA~2006 software \cite{ref28} are summarized in Table~\ref{tab:tab1} alongside with the parameters of the mineral nabokoite.

Synthetic nabokoite \nabok{} possesses the tetragonal symmetry with space group P4/ncc (D$_{4h}^8$). Oxidation states of constituting cations are K$^{+1}$Cu$^{+2}_7$(Te$^{+4}$O$_4$)(S$^{+6}$O$_4$)$_5$Cl. The crystal structure of nabokoite is shown in Fig.~\ref{fig:fig2-structure}. There are three positions for Cu$^{2+}$ ions. Two Cu1 ions within Cu1O$_5$Cl bipyramids and four Cu3 ions within Cu3O$_6$ octahedra form wavy square \kagome{} network, while Cu2 ions within Cu2O$_4$Cl pyramids decorate it. The coupling of Cu2 ions to SKL layers is fully frustrated in P4/ncc crystal structure in the presumption of prevailing antiferromagnetic exchange interaction within the SKL.

Structural data do not show traces of other phases in the samples obtained. However, magnetic measurements (see below) indicate presence of paramagnetic defects of unknown origin amounting to 3---4\% per copper ion.

\section{Experimental Results}
\subsection{Magnetization}
\begin{figure}[t]
  \centering
  \includegraphics[width=\figwidth]{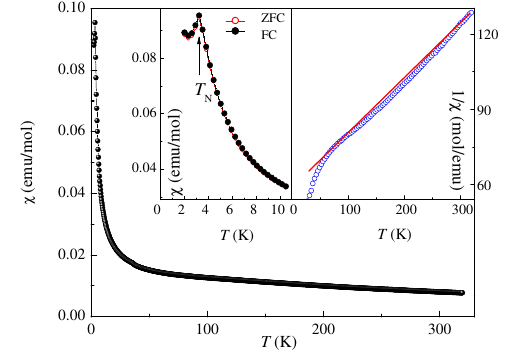}
  \caption{(color online) Temperature dependence of magnetic susceptibility in \nabok{} taken at $\mu_0 H = 0.1$~T. Left inset enlarges the low temperature region for both ZFC and FC regimes. Right inset represents the inverse magnetic susceptibility curve. Solid line is the guide for an eye. }\label{fig:fig3-chi}
\end{figure}

\begin{figure}[t]
  \centering
  \includegraphics[width=\figwidth]{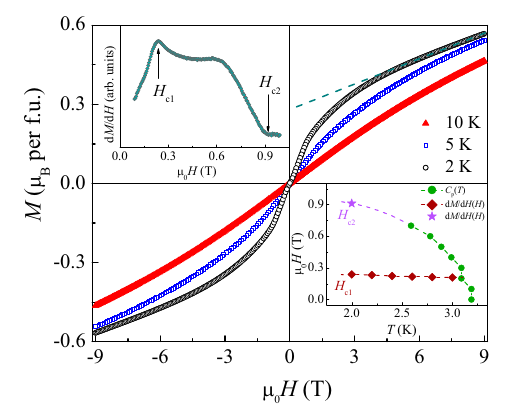}
  \caption{(color online) The field dependences of magnetization $M$ taken at various temperatures in \nabok{}. Dash line represents the linear extrapolation of the $M(H)$ curve at 2 K to zero magnetic field. Upper inset: the derivative $dM/dH$ of the curve taken at 2 K. Lower inset: magnetic phase diagram of \nabok{}. }\label{fig:fig4-M(H)}
\end{figure}

Thermodynamic properties of \nabok{}, i.e. magnetization $M$ and specific heat $\Cp$, were measured on the pressed pellets of the title compound using relevant options of the ``Quantum Design'' Physical Properties Measurements System PPMS-9T in the temperature range 2---320 K under magnetic field $\mu_0 H$ up to 9~T. The temperature dependence of the magnetic susceptibility $\chi = M/H$ taken in the field-cooled (FC) regime at $\mu_0 H = 0.1$~T is shown in Fig.~\ref{fig:fig3-chi}. The $\chi(T)$ curves taken in both FC and ZFC (zero-field-cooled) regimes evidence sharp peak at $\TN = 3.2$~K, as enlarged in the left inset to Fig.~\ref{fig:fig3-chi}.

Temperature dependence of the inverse magnetic susceptibility is shown in the right inset to Fig.~\ref{fig:fig3-chi}. In a wide temperature range, the $\chi^{-1}(T)$ curve is concave, meaning that a broad correlation hump is present at about 150~K in the $\chi(T)$ dependence. To underline the concavity, the solid line is drawn in the right inset to Fig.~\ref{fig:fig3-chi}.

The fitting of experimental data in the range 270--320 K by the Curie - Weiss law
\begin{equation}\label{eqn:curie}
\chi=\chi_0+\frac{C}{T-\Theta}
\end{equation}					

\noindent gives the temperature-independent term $\chi_0 = (8.8\pm 0.1) \times 10^{-4}$~emu/mol and the Weiss temperature $\Theta=-( 142.0\pm 0.6)$~K at the fixed value of the Curie constant $C = ng2S(S+1)/8 = 3.18$~emu$\cdot$K/mol, here  $n = 7$ is  the number of Cu$^{2+}$ ions,  $g = 2.2$ is a typical $g$-factor value and spin $S = 1/2$. Notably, the approximation with three independent parameters, $C$, $\Theta$ and $\chi_0$, gives an inadequate estimation of the effective magnetic moment, which points to the fact that the Curie-Weiss law is not strictly valid in this temperature range. The estimation of $\chi_0$ is positive and significantly exceeds the summation of Pascal's diamagnetic constants of constituent ions $\chi_\textrm{dia} = - 4.4 \times 10^{-4}$~emu/mol \cite{newref28}, while van Vleck's contribution for Cu$^{2+}$ in the cubic symmetry would be $\chi_\textrm{vV} = + 3 \times 10^{-4}$~emu/mol (for seven Cu$^{2+}$ ions per formula unit) \cite{newref29}. However, the last estimation is not quite suitable for nabokoite, since most copper ions in this compound are situated in a distorted mixed oxygen-chlorine environment. It should also be noted that the evaluations of $\Theta$ and $\chi_0$ parameters are not independent.

The field dependences of magnetization $M(H)$ in \nabok{} taken at various temperatures are shown in Fig.~\ref{fig:fig4-M(H)}. No hysteresis was seen in the magnetically ordered phase at $T < \TN$. The extrapolation of the $M(H)$ curve taken at 2 K by linear dependence results in residual magnetization of about $0.3\mu_B$  per formula unit which corresponds roughly to 4\% of impurities per copper ion of nabokoite. However the magnetic susceptibility $\chi$  below $\sim 50$~K  exceeds by far possible contribution from these impurities, meaning that the measured magnetization is due to copper spins of nabokoite.

The derivative $dM/dH$ at $T = 2$~K (see upper inset to Fig.~\ref{fig:fig4-M(H)}) reveals two characteristic fields of magnetization process. First, there is a well defined peak in $dM/dH$ at the field $\mu_0 H_{\rm c1}=0.23$~T, such peaks are typical for spin-flop transition in an antiferromagnet. Second, there is a change of slope of $dM/dH$ at $\mu_0 H_{\rm c2}=0.9$~T which is typical for a spin-flip transition in an antiferromagnet. In upper inset to Fig.~\ref{fig:fig4-M(H)}, the arrows mark the fields of the peak $H_{\rm c1}$ and sharp change of slope $H_{\rm c2}$.

\subsection{Specific heat}
\begin{figure}[t!]
  \centering
  \includegraphics[width=0.8\columnwidth]{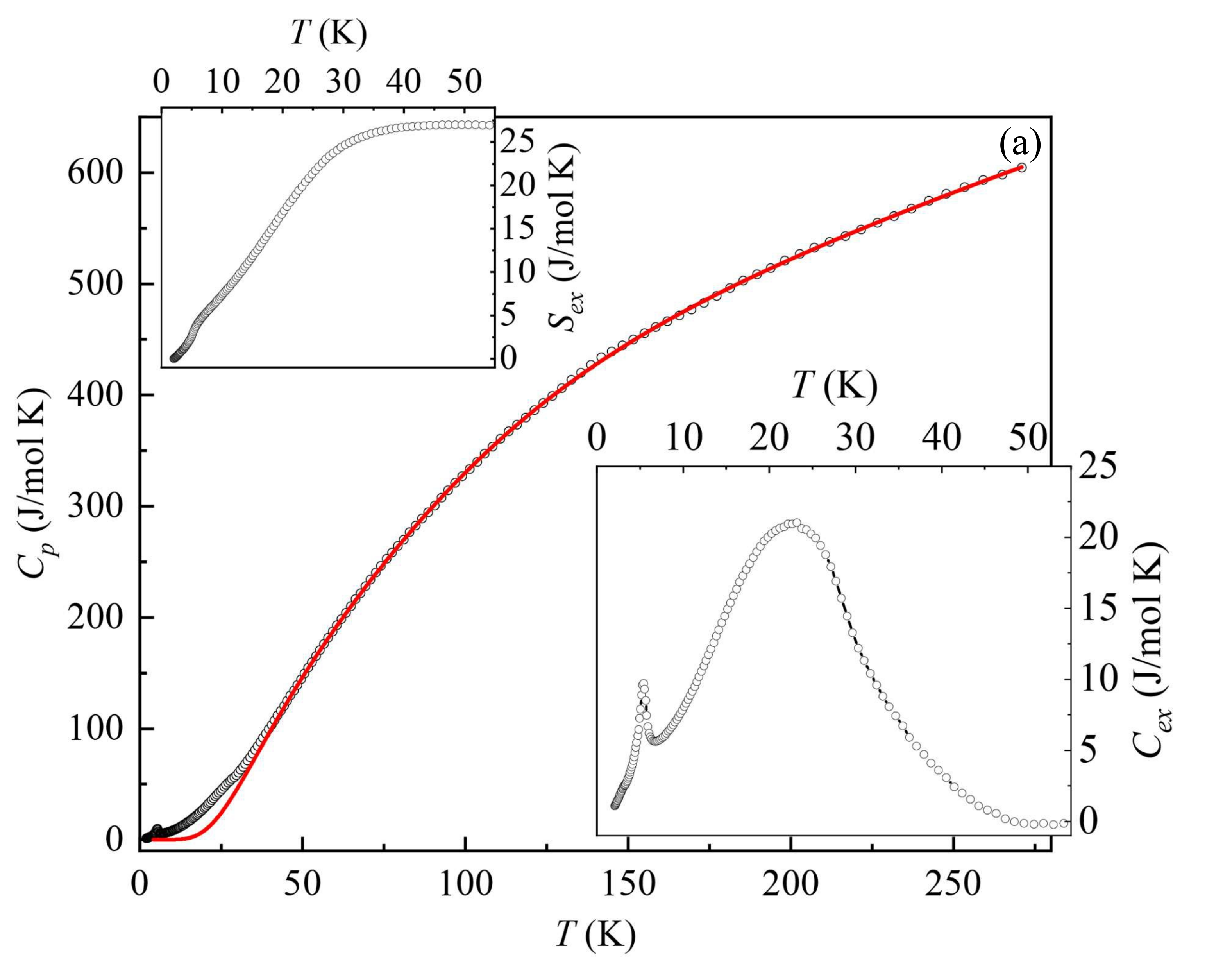}
  \includegraphics[width=0.48\columnwidth]{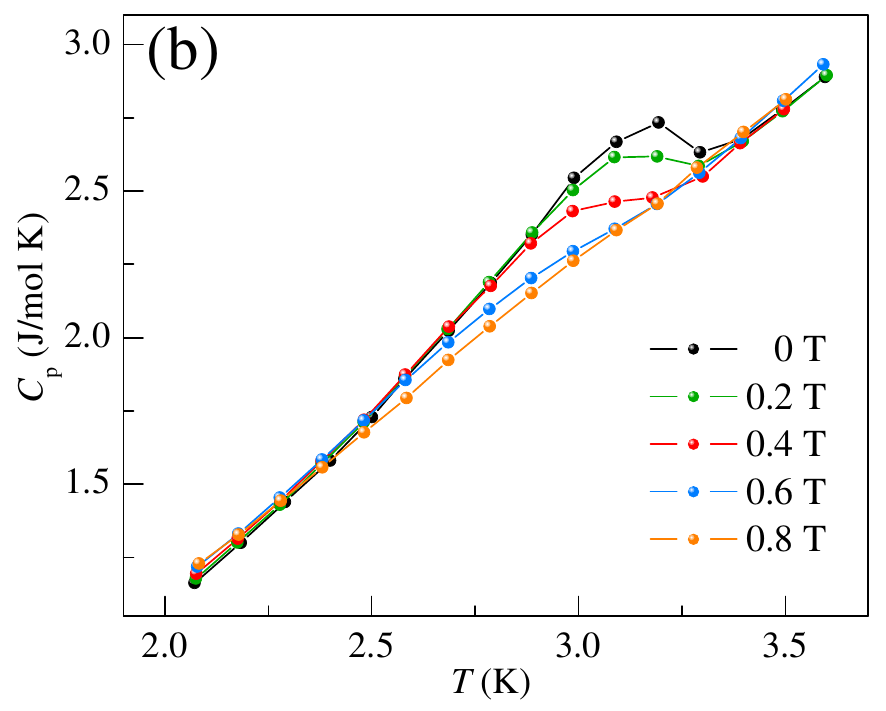}
  \includegraphics[width=0.48\columnwidth]{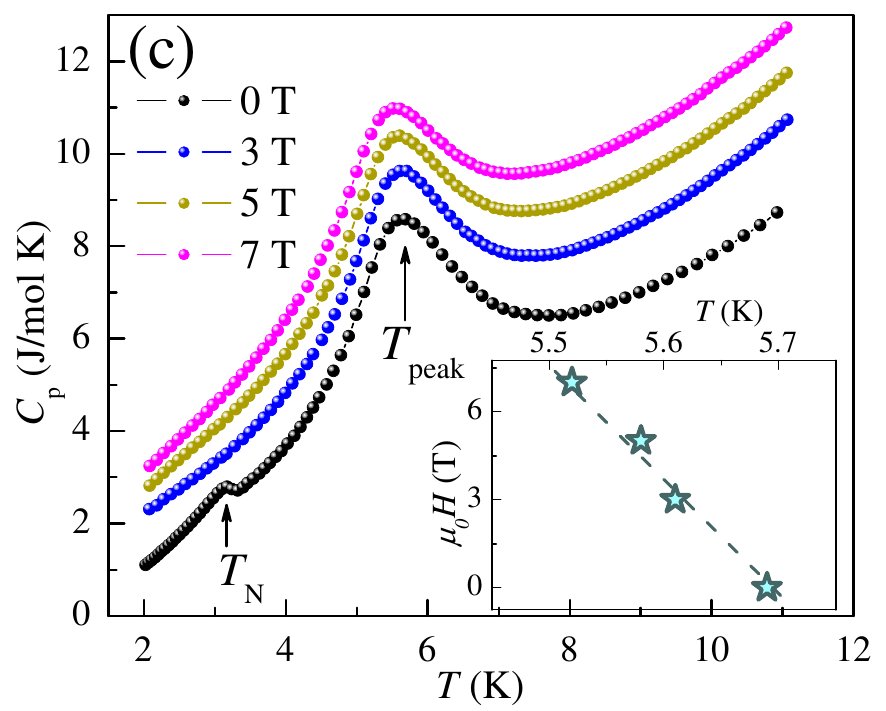}
  \caption{(color online) (a): Symbols --- temperature dependence of specific heat $\Cp$ in   \nabok{} taken at zero magnetic field, solid curve --- fitting of the experimental data by the sum of Debye and two Einstein functions as described in the text. Lower inset: excess specific heat $C_\textrm{ex}(T)$, upper inset: excess entropy $S_\textrm{ex}(T)$.   (b) Enlarged $\Cp(T)$ curves at $\mu_0 H < 1$~T in the N\'{e}el point vicinity. (c): Enlarged $\Cp(T)$ curves in the vicinity of $\Tpeak$  at $\mu_0 H = 0,~3,~5$ and 7~T, curves for $\mu_0 H\neq 0$ are shifted for clarity of representation.  Inset: magnetic field dependence of $\Tpeak$.}\label{fig:fig5-c}
\end{figure}

Two sharp anomalies are seen in $\Cp(T)$ dependence at $\TN = 3.2$~K and $\Tpeak = 5.7$~K, as shown in Fig.~\ref{fig:fig5-c}. The sharp anomaly at $\TN$ is suppressed by external magnetic field $\mu_0 H< 1$~T, as shown in the Fig.~\ref{fig:fig5-c}-b. Broad anomaly at $\Tpeak$ is quite robust to magnetic field slightly shifting to lower temperatures as field increases, as shown in Fig.\ref{fig:fig5-c}-c. Under magnetic field of 7 T, it shifts downward by less than 0.2~K. The shape of this anomaly differs from the standard shape of the Schottky anomaly being significantly narrower. Such a peak accompanied by a sharp low-temperature shoulder at $T = 0.04 J-0.1 J$, where $J$ is the leading exchange interaction parameter has been predicted for SKL  \cite{ref4,ref7}. Small sensitivity of the anomaly at $\Tpeak$ to external magnetic field allows to suggest its singlet-singlet origin.

The fitting of experimental $\Cp(T)$ dependence by the sum of Debye and two Einstein functions with the constrain on corresponding weights $\sum a_i = 39$ is shown by solid line in Fig.~\ref{fig:fig5-c}-a. The best fit is obtained at the Debye temperature $\Theta_\textrm{D} = 549$~K with $a_\textrm{D} = 15$ and the Einstein temperatures $\Theta_\textrm{E1} = 152$~K with $a_\textrm{E1} = 10$ and $\Theta_\textrm{E2} = 1453$~K with $a_\textrm{E2} = 14$. The residual between the experimental data and the fitting curve allows determining the excess specific heat $C_\textrm{ex}$, as shown in the lower inset to Fig.~\ref{fig:fig5-c}-a, and excess entropy $S_\textrm{ex}$, as shown in the upper inset to Fig.~\ref{fig:fig5-c}-a. Excess specific heat curve features  additional broad  anomaly at about 25~K reflecting  dielectric  permittivity anomaly described below. Evidently, very small portion of magnetic entropy releases below $\TN$ while the largest portion of the excess entropy is released at dielectric transition at about 25 K. Since the magnetic specific heat $C_\textrm{magn}$ is distributed over very wide temperature range, the excess entropy released at low temperatures is lower than the expected magnetic entropy $S_\textrm{magn} = nRln(2S + 1) = 40.3$~J/(mol$\cdot$K), here $n = 7$ is the number of magnetically active ions per formula unit and $R = 8.314$~J/(mol$\cdot$K) is the universal gas constant.

\subsection{Dielectric permittivity}
\begin{figure}[t]
  \centering
  \includegraphics[width=\columnwidth]{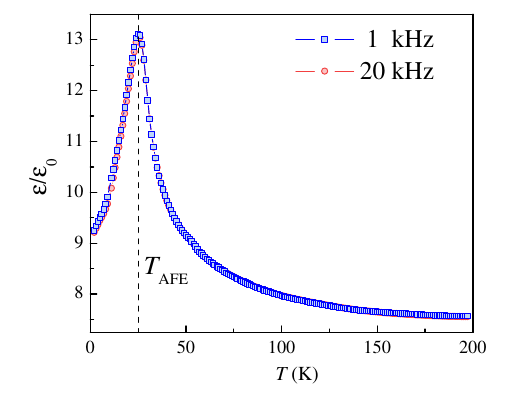}
  \caption{(color online) Temperature dependences of dielectric permittivity in \nabok{} at various frequencies. }\label{fig:fig6-epsilon}
\end{figure}

Temperature dependence of dielectric constant $\varepsilon$ in \nabok{} have been measured at various frequencies by a capacitance bridge Andeen-Hagerling 2700A on a thin pressed pellet covered by a silver paste. Well defined peak in $\varepsilon(T)$ curves is evidenced at $\Tafe = 25.4$~K, as shown in Fig.~\ref{fig:fig6-epsilon}. Its position and magnitude are independent on frequency in the range $f = 1 - 20$~kHz. The shape of this anomaly is appropriate for an antiferroelectric transition \cite{ref30,ref31}. Anomaly of ultra-high-frequency (tens of GHz range) dielectric constant at $\Tafe$ was also detected in ESR experiments.

While the nature of this transition is not defined yet, segnetoelectric transitions are usually accompanied by the lattice distortions. The lowering of the lattice symmetry can have two effects on the nabokoite spin system: (i) it could reduce the frustration of exchange interactions between decorating Cu2 ions and the square \kagome{} lattice of nabokoite; (ii) it could distort exchange bonds within the square \kagome{} layers. Note however, that the  basic properties of 2D square \kagome{} lattice remain intact even in the case of heavily distorted \kagome{} network \cite{ref12,ref15,ref16}. Analysis of magnetic resonance results (see Discussion) indicates that tetragonal symmetry is preserved at low temperatures.

\subsection{Electron spin resonance}
\begin{figure}[t!]
  \centering
  \includegraphics[width=\columnwidth]{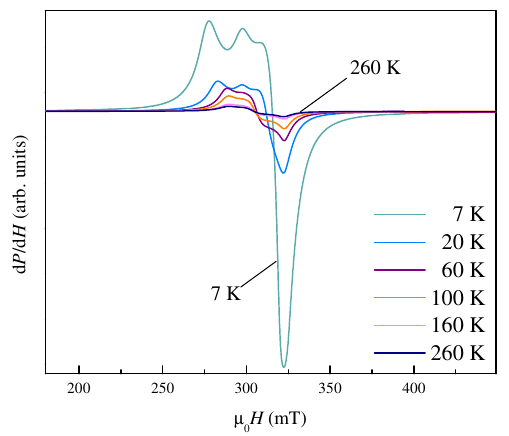}
  \includegraphics[width=\columnwidth]{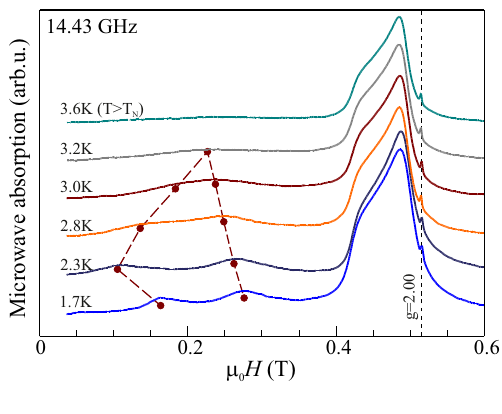}
  \caption{(color online) Representative spectra of paramagnetic ESR absorption in \nabok{} at selected temperatures. Top panel: X-band data from 7 to 260 K, field derivative of absorption is shown. Bottom panel: Transformation of ESR absorption at N\'{e}el point at $f = 14.43$~ GHz. Narrow peak at 0.515~T is a DPPH marker at g = 2.00. Filled circles connected with dashed line show position of antiferromagnetic resonance absorption. }\label{fig:fig7-spectra}
\end{figure}

\begin{figure}[t]
  \centering
  \includegraphics[width=\figwidth]{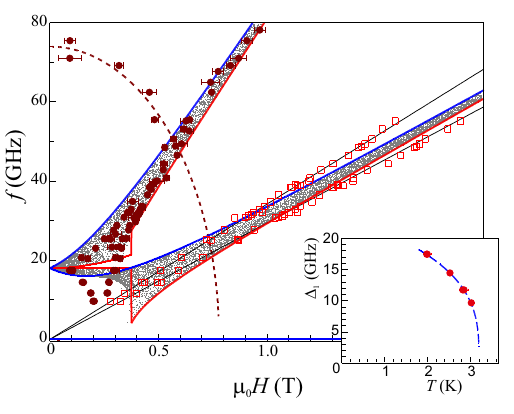}
  \caption{(color online) Frequency-field diagram for \nabok{} at 1.7 K. Squares mark the edges of powder-averaged paramagnetic absorption, thin straight lines corresponds to the boundaries of paramagnetic absorption with  $g = 2.44$ and $g = 2.10$. Circles mark the positions of maximums of antiferromagnetic resonance absorption. Solid curves --- modeled $f(H)$ for helicoidally ordered antiferromagnet  for $H||z$ and $H\perp z$, scattered small dots --- modeled $f(H)$ for powdered helicoidally ordered antiferromagnet (each dot corresponds to randomly chosen field direction), dashed curve --- phenomenological fit of the high-frequency mode. Inset: temperature dependence of the lower magnon gap measured in ``temperature resonance'' experiment. Symbols --- experiment, dashed line --- phenomenological fit with critical exponent. }\label{fig:fig8-f(H)}
\end{figure}

Electron spin resonance (ESR) spectra were obtained on the mixture of non-oriented single crystals of \nabok{} by means of X-band ESR spectrometer CMS 8400 Adani ($f=9.4$~GHz, $\mu_0 H =0 - 0.7$~T, $T = 7-300$~K) in Lomonosov Moscow State University and multi-frequency low-temperature ESR spectrometers ($f = 9-80$~GHz, $\mu_0 H = 6$~T, $T = 1.5-40$~K) in Kapitza Institute for Physical Problems. Representative ESR spectra at selected temperatures are shown in Fig.~\ref{fig:fig7-spectra}. Characteristic powder-averaged paramagnetic resonance spectra of the sample with anisotropic $g$-factor are observed from the lowest temperature of 1.5~K to room temperature. Edges of paramagnetic resonance absorption above 20~K correspond to the $g$-factor values of $2.33\pm 0.02$ at lower field edge and $2.07 \pm0.02$ at higher field edge. On cooling below 20~K, the lower field edge shifts to position corresponding to $g = 2.44 \pm 0.02$ (See Supplementary Information for details). Integral intensity of the ESR absorption in the vicinity of paramagnetic resonance fields follows Curie law from room temperature down to 1.5~K and does not demonstrate any anomaly at the N\'{e}el point. Comparison of the observed ESR intensity vs. ESR absorption in the fixed amount of the known paramagnet (see details in Supplementary Information) revealed that paramagnetic ESR response corresponds to $(3 \pm 1)$\% of $g = 2.2$ $ S = 1/2$ paramagnetic centers per Cu$^{2+}$ ion. This value of paramagnetic centers concentration agrees well with the estimation from $M(H)$ data analysis.

Thus we can conclude that small (3\%) amount of free spins responsible for the observed absorption signal and for low-temperature magnetization offset (see Fig.~\ref{fig:fig4-M(H)}) is decoupled from the main magnetic system of nabokoite. The main copper matrix of nabokoite does not contribute to the observed paramagnetic ESR absorption. Presumably, it is due to the strong relaxation processes in the copper matrix of nabokoite which broadens ESR line beyond limits of observation. Note, that quite large ESR linewidth of 1---2~T was also reported for \kagome{} compound herbertsmithite \cite{ref32}.

However, there are qualitative changes of ESR absorption below $\TN$: weaker components of absorption develop in the area of the smaller fields, as shown in the bottom panel of Fig.~\ref{fig:fig7-spectra}. The positions of these new maximums of absorption are temperature dependent, they disappear above $\TN$. The intensity of this antiferromagnetic resonance (AFMR) absorption signal can not be used as a measure of the quantity of the ordered phase since, as will be argued later, observed AFMR absorption corresponds to unusual non-Larmor resonance mode with unknown high-frequency susceptibility.

Additional absorption at $T \leq \TN$ was observed at the frequencies from 9 to 80~GHz. Resulting field-frequency diagram is shown in Fig.~\ref{fig:fig8-f(H)}. This diagram features two gaps at zero field $\Delta_1 = (18 \pm 1)$~GHz and $\Delta_2 = (75 \pm 3)$~GHz (values correspond to 1.7~K). Both gaps are temperature dependent. Within the mean field approach, gap in the spin waves spectrum is proportional to the antiferromagnetic order parameter (sublattice magnetization)  The temperature evolution of the lower gap $\Delta_1$ was followed in the zero-field ``temperature resonance'' experiment where the microwave power absorbed by the sample as a function of temperature at various frequencies was recorded. Once the frequency of the experiment coincides with the gap $\Delta_1$ the magnons are excited and the microwave power absorption is detected. Observed temperature dependence of the gap (see inset to Fig.~\ref{fig:fig8-f(H)}) can be fitted by phenomenological critical law $\Delta_1 = A(\TN-T)^\beta$ with N\'{e}el temperature fixed at 3.2~K and exponent $\beta\approx0.32$, this value is close to the values of order parameter critical exponents expected for a 3D antiferromagnet \cite{critexp1,critexp2}.

As we are dealing with powder sample (non-oriented tiny single crystals), we cannot ascribe positions of the maximums of absorption in powder-averaged ESR signal to some known modes of antiferromagnetic resonance. However, we can see that positions of these maximums form a regular pattern resembling a conventional softening of one of the antiferromagnetic resonance modes at spin-flop transition. Field of spin-flop transition estimated from the softening of this low-frequency "mode" is $(0.235 \pm 0.005)$~T in agreement with static magnetization data shown in the inset to Fig.~\ref{fig:fig4-M(H)}.

Magnetic resonance mode originating from the larger gap $\Delta_2$ decreases with the field and, being extrapolated to zero frequency value, probably softens at about $(0.75\pm0.10)$~T. This value is close to the field $H_{\rm c2}$, where magnetization derivative changes as shown in the inset to Fig.~\ref{fig:fig4-M(H)}.  However, we were not able to follow this mode at lower frequency and we observe no clear anomalies at the field of $0.6-1.0$~T. Also, we have to note that high-frequency antiferromagnetic resonance mode (at 70-80~GHz) is observed in the fields as high as $(0.96\pm0.02)$~T. Observed resonance mode approach linear asymptote with the slope 0.825~GHz/T, which would correspond to the $g$-value of 5.9 in the case of a paramagnetic resonance. This value of effective gyromagnetic ratio is unexpected for copper ions and, as will be discussed later, suggests that magnetic ordering in \nabok{} is non-collinear.

\section{Discussion}

\subsection{Basic features of SKL in nabokoite}
\begin{figure}[t]
  \centering
  \includegraphics[width=\figwidth]{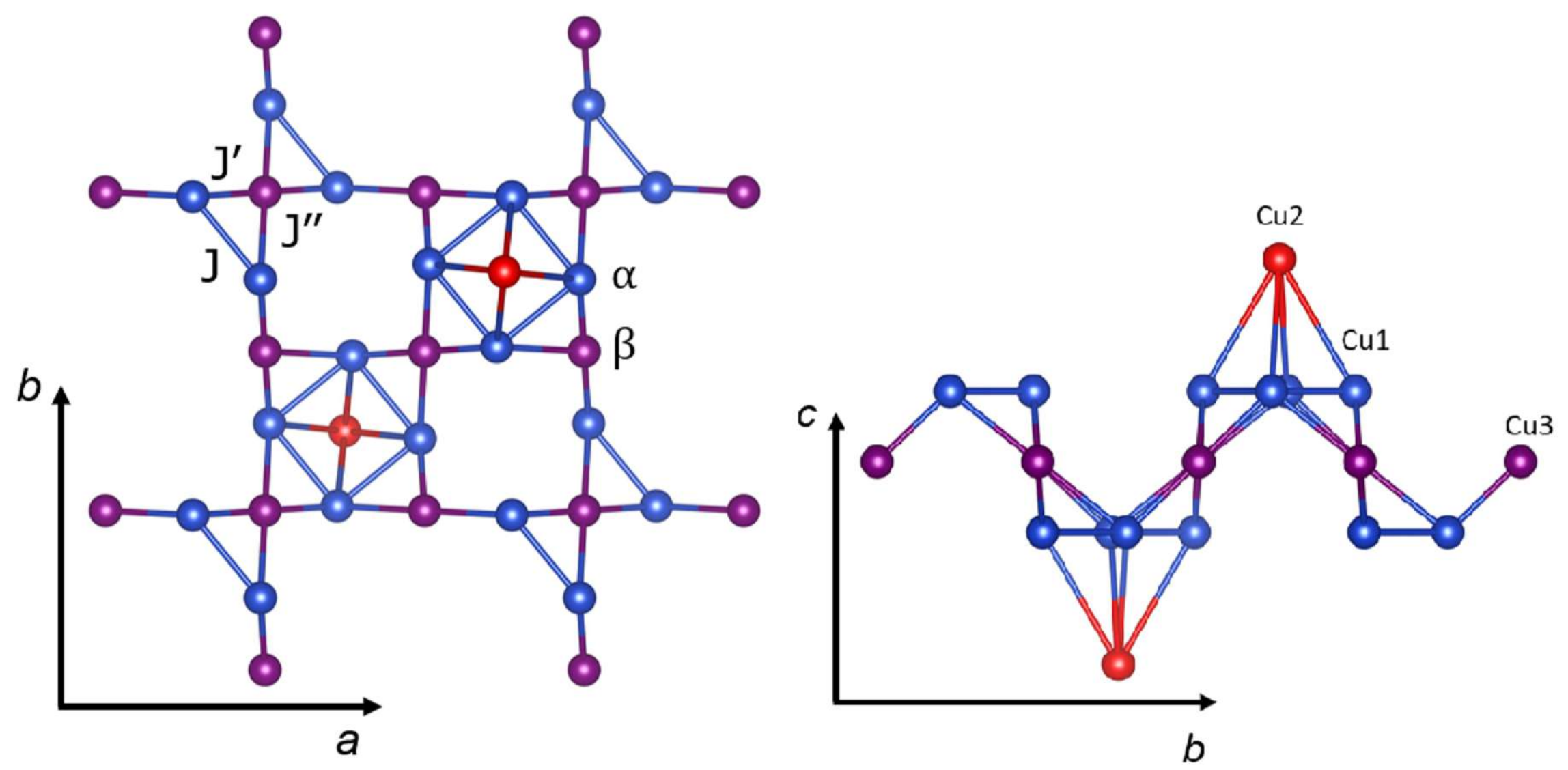}
  \caption{(color online) The magnetic network in \nabok{} viewed along the $c$ axis (left panel) and along the $a$ axis (right panel). }\label{fig:fig9-skl}
\end{figure}

The magnetic subsystem of \nabok{} is represented by the buckled SKL layers composed by Cu1 and Cu3 ions and decorated by interlayer Cu2 ions. The square \kagome{} lattice is a heavily distorted one with three inequivalent distances $(\alpha-\alpha) = 3.2943$~\AA{}, $(\alpha-\beta)' = 3.1022$~\AA{} and $(\alpha-\beta)'' = 3.4532$~\AA{}, which correspond to inequivalent exchange interaction pathways $J$, $J'$ and $J''$, as shown in Fig.~\ref{fig:fig9-skl}.

Initially, the theoretical treatment has been given to equilateral SKL presuming that all exchange interaction parameters between $\alpha$  and $\beta$   sites are equal. At $x = J_{\alpha\beta}  /J_{\alpha\alpha}   = 1$, SKL represents a system of equilateral corner-sharing triangles, which can be treated within the resonating valence bond scenario \cite{ref9,ref10,ref13}. The main results of this consideration were the prediction of a multitude of low-lying singlet states, a singlet-triplet gap and a characteristic temperature dependence of a specific heat featuring two peaks \cite{ref1,ref4,ref7}.

An important step towards experimentally realized lattices was an assumption of $J_{\alpha\beta}\neq J_{\alpha\alpha}$. At $x\gg 1$, SKL becomes essentially bipartite forming a collinear ferrimagnetic ground state, say $\alpha$ spins up and $\beta$ spins down. In the opposite limit, $x \ll 1$, the $\alpha$ spins form isolated singlet tetramers alongside with free dangling $\beta$ spins.  Two types of competing valence bond crystal (VBC) phases were found in the vicinity of $x = 1$. At $x<1$ and $1 < x < 1.75$, the pinwheel VBC states were predicted, while the loop-six VBC state was predicted for $x > 1.75$  \cite{ref10}.

The case of isosceles SKL with $J_{\alpha\beta}\neq J_{\alpha\alpha}$ corresponds to Na$_6$Cu$_7$BiO$_4$(PO$_4$)$_4$Cl$_3$ which, however, shows neither a singlet-triplet gap nor a phase transition to ferrimagnetic state down to ultralow temperatures \cite{ref27}. The magnetic phase diagram of SKL magnet with three nonequivalent exchange interactions has been analyzed also \cite{ref16}. Depending on ratios of these three exchanges numerous ground states were predicted including spin liquid, N\'{e}el order and up-up-down configuration. The case of $J_{\alpha\alpha}\neq J_{\alpha\beta'}\neq J_{\alpha\beta''}$ corresponds to KCu$_6$AlBiO$_4$(SO$_4$)$_5$Cl which shows properties of gapless spin liquid down to ultralow temperatures \cite{ref25}.

Predicted temperature dependence of the specific heat for an equilateral SKL layer \cite{ref4,ref7} has two characteristic features. First, there is a broad hump on $\Cp(T)$ dependence located at $T\sim 0.67 J$ \cite{ref4} (to be compared with earlier estimation $T\sim 0.85 J$ \cite{ref7}). This feature, however, cannot be picked out in $\Cp(T)$ curve due to prevailing phonon contribution at elevated temperatures. Numerical results for $\chi(T)$ \cite{ref4} predicts also a weakly developed convex hump on $\chi(T)$ curve at $T=0.935 J$, similar behavior has been observed in our measurements on nabokoite at  150~K(see Fig.~\ref{fig:fig3-chi}), which gives estimation for the leading exchange interaction parameter $J \sim  160$~K. Second feature of $\Cp(T)$ dependence is a weak peak accompanied by a sharp low-temperature shoulder at $T\simeq 0.04J$---$0.1J$ \cite{ref4,ref7} attributed to low-lying singlet excitations filling the singlet-triplet gap.
The peak observed in specific heat at $\Tpeak = 5.7$~K can be identified with this prediction, its robustness with respect to magnetic field is naturally explained by singlet nature of relevant excitations. In this case, the estimation for the leading exchange interaction parameter lies in the limits $J \sim 55 -  140$~K.

\subsection{Low temperature phases in nabokoite}

We observed  two phase transitions at cooling of \nabok{}: there is a dielectric  anomaly at $\Tafe = 25.4$~K, which is followed by an antiferromagnetic ordering at $\TN = 3.2$~K. To exclude possible misinterpretation of the effects caused by these transitions on the physics of quantum ground state formed in SKL layers, we demonstrate below that our experimental results allows drawing a border between these ordered phases and properties of SKL layers of nabokoite.

Dielectric anomaly is a typical signature of ferroelectric phenomena, formation of ferro- or antiferroelectric order is usually accompanied by a lattice distortion. As was discussed above, nabokoite structure even at room temperature does not correspond to equilateral SKL model and include three nonequivalent exchange paths. Thus, small lattice distortions within the layer are not expected to drastically change this picture. As will be demonstrated below, analysis of magnetic resonance data in the antiferromagnetically ordered phase evidences that tetragonal lattice symmetry is actually conserved at low temperatures. Lattice symmetry lowering due to ferroelectric transition can have stronger effect on the coupling of the interlayer Cu2 ions to the rest of the nabokoite spin system. At room temperature, all Cu2-Cu1 paths are equivalent which results in total decoupling of Cu2 spins within mean field approximation. Lattice distortion can lift this frustration opening route to 3D magnetic ordering.

\nabok{} orders antiferromagnetically at $\TN = 3.2$~K, the N\'{e}el temperature is significantly lower than the characteristic scale of the exchange interaction parameters estimated in the thermodynamic measurement. Such ``freezing'' of a spin liquid, being driven not by the main strong exchange coupling but by some weaker interaction, results sometimes in complicated forms of magnetic ordering \cite{ref33,ref34}. Our experimental data allows to draw certain conclusions on the antiferromagnetic structure of nabokoite. These conclusions are based on the following experimental findings: (i) only few percent of magnetic entropy freeze out on cooling below the N\'{e}el point, (ii) low temperature electron spin resonance experiment reveals two gaps at zero field and (iii) one of the resonance modes below $\TN$ has asymptotic slope of $f (H)$ dependence equal to 0.825~GHz/T, which is strongly different from Larmor precession gyromagnetic ratio 0.280~GHz/T for $g = 2.00$ (see Fig.~\ref{fig:fig8-f(H)}).

Our present experiments cannot discern whether ordered state take over all copper ions of nabokoite, or only part of the magnetic ions orders below $\TN$.  This question can be answered in further NMR or neutron diffraction study. Since magnetic ordering temperature is almost twofold smaller than the temperature of the specific heat peak $\Tpeak$ both of the above scenarios does not contradict to the formation of the spin-liquid state within the SKL layers of \nabok{}. However, keeping in mind very small residual magnetic entropy at $\TN$ we presume that only the inter-planar Cu2 ions coupled through the virtual excitations from the spin liquid state form the antiferromagnet.

Observation of the $f(H)$ asymptotic slope almost three-fold different from the Larmor gyromagnetic ratio is not compatible with the well-established theory of antiferromagnetic resonance in collinear antiferromagnets \cite{ref35,ref36}. On the other hand, non-collinear antiferromagnets are known to demonstrate non-Larmor asymptotic slope for some of the resonance modes \cite{ref37}.

The presence of the two gaps in magnon spectrum is also important. Gaps can originate from anisotropic spin-spin interactions that make some orientations of the order parameter preferable thus giving rise to finite energy cost of low-frequency order parameter oscillations. Alternatively, the gaps can be of exchange origin corresponding then to the high-energy oscillations of the spin structure. The former case (gaps of anisotropic origin) can be treated within hydrodynamic approach \cite{ref38}. For a helix antiferromagnet, this model predicts single gap in the axial case and two gaps if the axial symmetry is violated. To describe our data, we fit them by analytical equations following \cite{ref37} and by numerical modeling taking into account randomly oriented powder particles using approach of \cite{ref39}. We found, that models with two gaps (which implies either lowering of the lattice symmetry at least down to  orthorhombic  or accounting for 4-th order anisotropy in the case of tetragonal lattice symmetry) disagree qualitatively with the experiment. All these models predict that the mode with non-Larmor slope should originate from the higher gap (see Supplementary Materials for details). Thus, we conclude that magnetic resonance mode originating from the higher gap $\Delta_2 =( 75\pm 3)$~GHz is of exchange origin and, hence, its tentative softening indicates transition with change (or total suppression) of antiferromagnetic order. We fit its field dependence by the phenomenological function $f=\Delta_2\sqrt{1-(H/H_{\rm c2})^2}$ with $\mu_0 H_{\rm c2}=0.78$~T, mimicking known result of molecular field theory for collinear antiferromagnet \cite{ref35}. This estimation of the critical field $H_{\rm c2}$ agrees with the magnetization data, shown in the upper inset to Fig.~\ref{fig:fig4-M(H)}.

The antiferromagnetic resonance data, as shown in Fig.~\ref{fig:fig8-f(H)}, are well described in the model assuming axial anisotropy (see \cite{ref37} and Supplementary Materials) of the helically ordered antiferromagnetic phase. Thus, we can ascertain that antiferroelectric transition at $\Tafe = 25.4$~K does not result in significant lowering of lattice symmetry. Besides of the gap value $\Delta_1$, this model depends  on the ratio of magnetic susceptibilities for the field applied along the vector normal to the helix plane $\chi_{||}$ and for the field applied orthogonally to this vector $\chi_\perp$. Ratio of these susceptibilities determines non-Larmor slope of one of the magnetic resonance modes, best fit of our data corresponds to $\chi_{||}/\chi_\perp =3.7$. Spin-flop field $\mu_0 H_\textrm{c1}$ obtained within model calculations is 0.36~T, somewhat higher than the experimentally found value. This difference could be the consequence of the helix deformation as the magnetic field is applied.

Summing up, while both antiferroelectric transition and magnetic ordering in nabokoite are subjects of separate interest, these phase transitions do not affect largely the properties of square \kagome{} lattice layers formed by Cu1 and Cu3 ions. Magnetic ordering in interlayer Cu2 subsystem probably coexists with spin liquid state within the layers.

\section{Conclusion}
We prepared the synthetic nabokoite, \nabok{}, and observed peculiar behavior of its square \kagome{} magnetic system decorated by dangling copper ions. Anomaly in specific heat at $\Tpeak = 5.7$~ K robust to external magnetic field can be attributed to transitions between various spin-singlet configurations in the ground state of SKL. The long-range antiferromagnetic state formation has been observed at $\TN = 3.2$~K, which can be attributed to the ordering of decorating Cu2 ions being coupled through virtual excitations of spins from the network of Cu1 and Cu3 ions in the ground state of SKL. Magnetic resonance data indicate that magnetic order in nabokoite is non-collinear. Additionally, evidences were obtained on the formation of antiferroelectric order at low temperatures, which deserves a separate study. Compared to theoretical predictions, the intrinsic SKL physics is largely masked in real compounds.

\acknowledgements

We acknowledge useful discussions with S.~Li, L.~Shvanskaya, S.~Streltsov and J.~Richter. The synthesis of the samples was supported by Russian Science Foundation grant 23-23-00205. The X-ray data collection was performed on the equipment supported by the Lomonosov Moscow State University development program. Thermodynamic measurements were supported by the Megagrant program of Russian Government through the project 075-15-2021-604.  Electron spin resonance measurements in P.~Kapitza Institute for Physical Problems were supported by Russian Science Foundation grant 22-12-00259.

\section*{Supplementary Information:  details of ESR experiments}
\appendix
\renewcommand{\appendixname}{Supplementary Information}

\section{ESR experimental details summary}
Electron spin resonance (ESR) spectra were obtained on the mixture of non-oriented single crystals of \nabok{} by means of X-band ESR spectrometer CMS 8400 Adani ($f=9.4$~GHz, $\mu_0 H =0 ... 0.7$~T, $T = 7...390$~K) in Lomonosov Moscow State University and multi-frequency low-temperature ESR spectrometers ($f = 9...80$~GHz, $\mu_0 H = 6$~T, $T = 1.5...40$~K) in Kapitza Institute for Physical Problems.

X-band spectrometer CMS 8400 Adani uses conventional field modulation technique and measures field derivative of the absorption $d P/d H$. Multi-frequency ESR spectrometers in Kapitza Institute measure microwave power transmitted through the cavity with the sample, decrease of the power transmitted is the microwave power absorbed by the sample.

\section{Paramagnetic resonance line shape analysis}

\begin{figure}
  \centering
    \includegraphics[width=\figwidth]{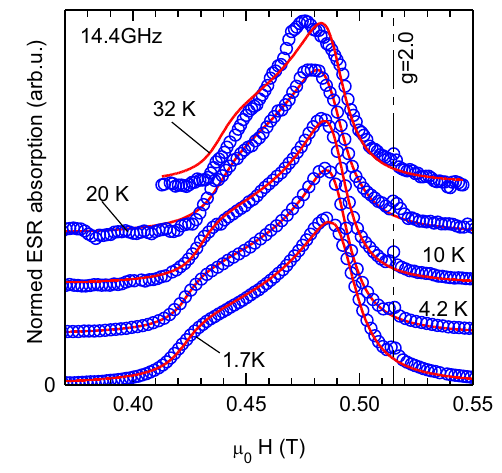}
 \caption{(color online) Symbols --- ESR absorption at different temperatures normed by its' maximum value at given temperature, solid curves --- best fit in a model of powder averaged ESR line with axial $g$-tensor and isotropic linewidth.}
\label{SM:fig:fig1-spectra}
\end{figure}

\begin{figure}
  \centering
 \includegraphics[width=\figwidth]{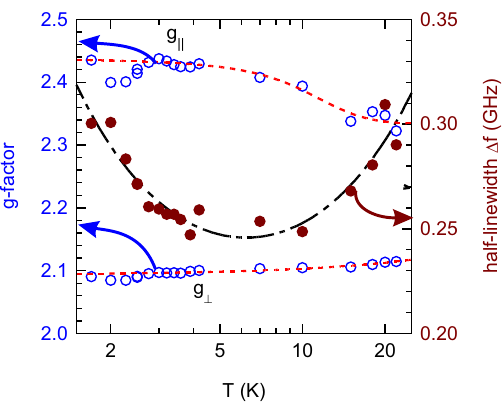}
 \caption{(color online) Temperature dependencies of $g$-tensor main values $g_{||}$ and $g_\perp$ (left Y-scale) and model half-linewidth $\Delta f$ (right Y-scale) for the powder model as determined from the fit of the data shown at Fig.~\ref{SM:fig:fig1-spectra}.}
\label{SM:fig:fig2-g(T)}
\end{figure}

As described in the main manuscript, shape of the paramagnetic resonance  absorption line in nabokoite samples looks like typical powder-averaged ESR signal. We model ESR absorption  numerically by averaging over powder particles orientation. We assume powder particles to be equally distributed over all orientation. Assuming conventional transverse polarization of microwave radiation, axial $g$-tensor and Lorentzian line shape with isotropic linewidth $\Delta f$ in frequency domain one can write for the power absorbed at a given field  $H$:

\begin{equation}
\label{SM:eqn:PMlineshape}
P(H)\propto \int_0^{\pi/2}\frac{[g_\perp^2+\frac{1}{2}(g_{||}^2-g_\perp^2)\sin^2\Theta] \sin\Theta}{1+((f-g(\Theta)\mu_B H/h)/\Delta f)^2}d\Theta
\end{equation}

\noindent here $\Theta$ is a polar angle, $h$ - Planck constant, $f$ - microwave frequency of ESR experiment, $\Delta f$ - linewidth in frequency domain and effective $g$-factor $g(\Theta)=\sqrt{g_{||}^2 \cos^2\Theta+g_\perp^2 \sin^2\Theta}$, bracheted expression in the numerator is the result of averaging of squared effective $g$-factor along the microwave field over all possible powder particles with given $\Theta$. Field derivative required for simulation of X-band EPR spectra can be directly deduced from this equation.

We tried to fit experimental data with Eqn.~(\ref{SM:eqn:PMlineshape}) for high-frequency ESR experiments or its derivative for X-band experiments. Fit routine was realized with Octave \cite{octave} software with its standard minimization routines. We did not succeeded to fit paramagnetic absorption in full temperature range ($1.7 ... 300$~K) by a combination of two powder-averaged signals, the fit procedure converges badly and is not stable. However, low-temperature high-frequency data below 20~K can be reasonably approximated by a single powder-averaged component with $g_{||}>g_\perp$. Above 20~K this fit becomes unsatisfactory indicating presence of at least two powder-averaged signals (see Fig.~\ref{SM:fig:fig1-spectra}).

Temperature dependencies of  $g$-tensor main values and model linewidth determined in these simulations are shown at Fig.~\ref{SM:fig:fig2-g(T)}. Transverse $g$-factor component remains almost constant $g_\perp=(2.07\pm0.02)$, while longitudinal component $g_{||}$ smoothly changes from $(2.33\pm0.02)$ at 20~K to $(2.44\pm0.02)$ at 1.7~K. Model linewidth changes slightly with temperature, increase of the model linewidth with heating above 10~K is most likely due to the presence of unresolved spectral components, since the fit quality decreases with heating (see Fig.~\ref{SM:fig:fig1-spectra}).

\section{Details of absolute calibration of ESR absorption}

\begin{figure}
  \centering
 \includegraphics[width=\figwidth]{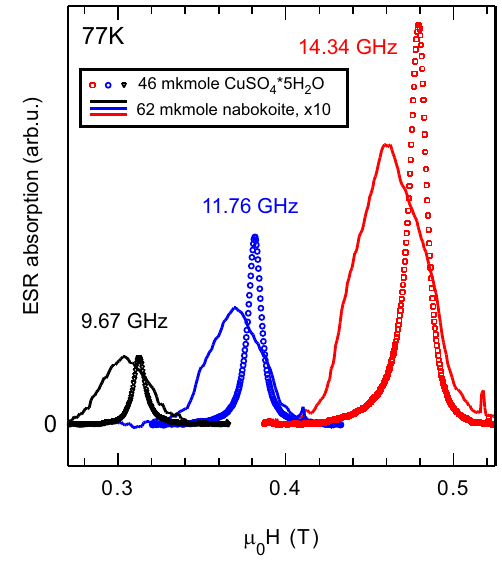}
 \caption{(color online) Comparison of ESR absorption for 73.7~mg (62~mkmole) nabokoite sample (curves, data are Y-amplified by a factor of 10) and 11.6~mg (46~mkmole) \cuso{} sample (symbols). $T=77$~K.}
\label{SM:fig:fig3-calibrate}
\end{figure}

\begin{figure*}
  \centering
  \includegraphics[width=\textwidth]{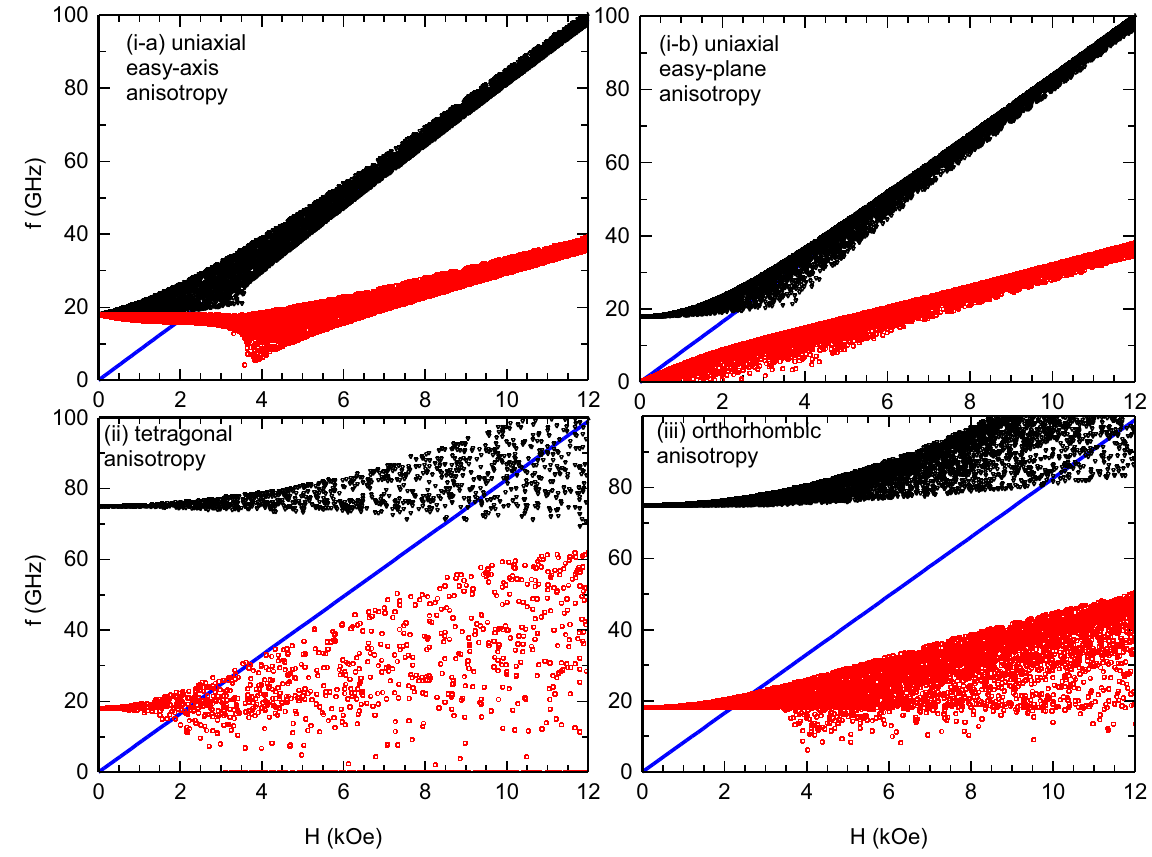}
 \caption{(color online) Comparison of the result of $f(H)$ modeling for powdered helicoidal antiferromagnet within hydrodynamic approach for different forms of anisotropy energy (see text). Each point corresponds to random field direction. Model parameters are tuned to reproduce smaller zero field gap for uniaxial form of anisotropy and both gaps for orthorhombic and tetragonal models, as well as to reproduce asymptotic slopes of  resonance modes. Straight line corresponds to the observed slope of 8.25~GHz/kOe.}
\label{SM:fig:f(H)}
\end{figure*}

We used two approaches for absolute calibration of ESR absorption intensity. Data from multi-frequency ESR spectrometers (Kapitza Institute) were used for calibration.

First, we can calculate imaginary part of high-frequency magnetic susceptibility. We did this calibration for 4.2...6K data at 14.43 GHz. Voltage drop on the quadratic detector of the transmission type spectrometer is (see, e.g., \cite{pool}):

\begin{equation}
\label{SM:eqn:dU/U}
\frac{\Delta U}{U}=\frac{1}{(1+4\pi Q \eta \chi'')^2}
\end{equation}

\noindent here $Q$ is cavity Q-factor, which  was measured directly during the ESR experiment ($Q=2300\pm100$), filling factor $\eta$ is the ratio of microwave field energies within the sample and within the cavity and   can be estimated from the sample mass (in our case $\eta=6.9\cdot 10^{-3}$). Imaginary part of susceptibility is then recalculated to static susceptibility per unit volume via Kramers-Kroenig equation:

\begin{equation}
\label{SM:eqn:KK}
\chi(0)=\frac{2}{\pi} \int_0^\infty \frac{\chi''(\omega)}{\omega} d\omega \approx \frac{2}{\pi H_0}\int_0^\infty \chi''(H) d H,
\end{equation}
the last approximation corresponds to the case of paramagnet with narrow absorption line, $H_0$ being the mean resonance field. Finally, susceptibility per unit volume is recalculated to susceptibility per mole yielding $\chi(\textrm{4.2~K})=18\cdot 10^{-3}$~emu/mole and $\chi(\textrm{6.0~K})=11\cdot 10^{-3}$~emu/mole. Both values correspond approximately to 3\% of paramagnetic defects per copper ion of nabokoite.  Estimated uncertainties are $\sim 30$~\%, being mostly due to the uncertainty in filling factor calculation for the real sample.

Second approach is based on comparison of ESR integral absorption measured in the same spectrometer in nabokoite and in paramagnetic \cuso{}. We used 73.7~mg nabokoite sample  and 11.6~mg \cuso{}. Comparative measurements were done at three microwave frequencies at room temperature and at 77~K. Measured ESR absorption is shown at Fig.~\ref{SM:fig:fig3-calibrate}.Q-factor of the microwave cavity was measured  in each experiment to compensate for its possible change during sample replacement.

Integrated intensity of ESR absorption for  73.7~mg (62~mkmole) nabokoite sample amounts to $(27\pm4)$\% of integrated intensity of 11.6~mg (46~mkmole) \cuso{} sample. Neglecting $g$-factor difference (which is within accuracy of our estimate) this means that the observed  high-temperature ESR response of nabokoite corresponds to approx. 0.2 copper spin per molecule or 3\% of paramagnetic defects per copper ion of nabokoite.  Estimated uncertainties are $\sim 30$\%, main source of uncertainty is the neglected distribution of microwave field in the larger nabokoite sample.

Thus, both approaches agrees that intensity of the observed high-temperature paramagnetic response of nabokoite corresponds to $(3\pm 1)$\% of paramagnetic defects per copper ion.

\section{Details of antiferromagnetic resonance description}

We describe here details of calculations of $f(H)$ dependencies for antiferromagnetic resonance modes in nabokoite. As described in the main manuscript, observation of the resonance mode with the asymptotic slope 0.825~GHz/T (8.25~GHz/kOe) instead of conventional Larmor gyromagnetic ratio $\approx0.280$~GHz/T is in total disagreement with the known theory of magnetic resonance in the ``traditional'' two-sublattices antiferromagnets \cite{ref36,ref35}. On the other hand, such non-Larmor modes are known in non-collinear antiferromagnets.

Since detailed information on the magnetic order in \nabok{} is lacking, we assume mathematically simplest case of helicoidal non-collinear ordering in nabokoite. Hydrodynamic approach to the problem of antiferromagnetic resonance of non-collinear antiferromagnet \cite{ref38} describes spin helix by two orthogonal unit vectors $\vect{l}_1$ and $\vect{l}_2$, ordered spin component at $\vect{r}_i$  position being
  $$ \vect{S}_i=S(\vect{l}_1 \cos(\vect{q}\vect{r}_i+\varphi)+\vect{l}_2 \sin(\vect{q}\vect{r}_i+\varphi),$$
\noindent here $\vect{q}$ is spin helix propagation vector, which is assumed to be incommensurate with the lattice.

Low energy dynamics of such a helicoidal antiferromagnet can be described with Lagrangian density \cite{ref37}
\begin{eqnarray}
{\cal L}&=&\frac{I}{2}(\dot{\vect{l}}_1+\gamma[\vect{l}_1\times H])^2+\frac{I}{2}(\dot{\vect{l}}_2+\gamma[\vect{l}_2\times H])^2+{}\nonumber\\
&&{}+\frac{I_n}{2}(\dot{\vect{n}}+\gamma[\vect{n}\times H])^2-U_a\label{SM:eqn:L}
\end{eqnarray}

\noindent here $U_a$ is the anisotropy energy which depends on lattice symmetry, vector $\vect{n}=[\vect{l}_1\times\vect{l}_2]$ is the normal to the plane containing all spin vectors of helical structure, and constants $I$ and $I_n$ determine magnetic susceptibility tensor of the antiferromagnet. Magnetic susceptibility for the field applied parallel to the normal $\vect{n}$ is $\chi_{||}= 2 I \gamma^2$, and the susceptibility for the field applied orthogonally to the normal $\vect {n}$  is $\chi_\perp=(I+I_n)\gamma^2$, $\gamma$ being the gyromagnetic ratio. Usually $\chi_{||}>\chi_\perp$ and at high field plane of the helical spin structure should be orthogonal to the field. As we are not interested to reproduce exact values of susceptibilities, we set $I=1$ for simplicity.

Anisotropy energy depends on the lattice symmetry and on the orientation of the spin structure with respect to the lattice. In the case of spiral structure anisotropy energy depends only on the components of $\vect{n}$-vector due to the symmetry to arbitrary translation along the helix propagation vector. Dielectric constant anomaly observed in nabokoite at 25~K indicates possible lowering of the lattice symmetry with cooling. We have considered three possibilities: (i) axial anisotropy of the ordered phase $U_{a1}=\frac{A}{2} n_z^2$; (ii) tetragonal (4-th order) anisotropy $U_{a2}=\frac{A}{2} n_z^2+\frac{\alpha}{2}n_x^2 n_y^2$ and (iii) orthorhombic anisotropy $U_{a3}=\frac{A}{2}n_z^2+\frac{B}{2} n_x^2$. Note, that case (iii) effectively includes all possible symmetries below orthorhombic: for monoclinic or triclinic lattice symmetry some (or all) anisotropy axes became arbitrary oriented with respect to the lattice, but powder averaging makes this insignificant.
 \begin{table}
 \label{SM:tab:tab1}
 \caption{Zero-field eigenfrequencies  for different forms of anisotropy energy.}
 \begin{ruledtabular}
 \begin{tabular}{c|c|c}
Model& Anisotropy & Zero-field\\
&energy&eigenfrequencies\\
\hline
 (i-a) axial & $U_{a1}=\frac{A}{2} n_z^2$&$\omega_1=\omega_2=\sqrt{\frac{\abs{A}}{I+I_n}}$\\
 easy-axis &$A<0$&$\omega_3=0$\\
 \hline
 (i-b) axial & $U_{a1}=\frac{A}{2} n_z^2$&$\omega_1=\sqrt{\frac{\abs{A}}{I+I_n}}$\\
 easy-plane &$A>0$&$\omega_2=\omega_3=0$\\
 \hline
 (ii) tetragonal&$U_{a2}=\frac{A}{2} n_z^2+\frac{\alpha}{2}n_x^2 n_y^2$&$\omega_1=\sqrt{\frac{\abs{A}}{I+I_n}}$\\
 &$A<0$, $\alpha>0$&$\omega_2=\sqrt{\frac{\alpha}{I+I_n}}$\\
 &&$\omega_3=0$\\
 \hline
 (iii) ortho-&$U_{a3}=\frac{A}{2}n_z^2+\frac{B}{2} n_x^2$&$\omega_1=\sqrt{\frac{\abs{A}}{I+I_n}}$\\
 rhombic&$A<B<0$&$\omega_2=\sqrt{\frac{B-A}{I+I_n}}$\\
 (or below)&&$\omega_3=0$\\
 \end{tabular}
 \end{ruledtabular}
 \end{table}

Oscillations  eigenfrequencies can be found from the dynamics equations derived from Lagrangian (\ref{SM:eqn:L}) by variational technique or by use of Euler-Lagrange equations. High-field asymptotes of AFMR modes are $$\omega=\frac{I\pm I_n}{I+I_n}\gamma H,$$ observed slope corresponds to $I_n=-0.456$.  Results for the orthorhombic anisotropy   are known \cite{ref37}. Zero field eigenfrequencies are for all studied models are listed in the Table~\ref{SM:tab:tab1}.
Zero eigenfrequency $\omega_3$ is a feature of spiral ordering, it is due to particular degeneracy of the incommensurate helix towards arbitrary translation. Models (ii) and (iii) predicts two different magnonic gaps at zero field, all of these models predicts spin-reorientation transitions at certain fields.

To compare these models with the powder sample experiment (see Fig.~8 of the main manuscript) we have performed numerical calculation of the eigenfrequencies for the randomly distributed field directions. Calculations were performed using approach developed in \cite{ref39}. Anisotropy parameters were tuned to reproduce observed zero-field gaps $\Delta_1=18$~GHz and $\Delta_2=75$~GHz (only the smallest gap $\Delta_1$ was considered for axial model (i)), gyromagnetic ratio was chosen to correspond to $g=2.2$ ($\gamma=19.35 \frac{10^9 \textrm{ rad}}{\textrm{s}\cdot\textrm{kOe}}$) and $I=1$ and $I_n=-0.456$ as described above. Results of the calculations are shown at Fig.~\ref{SM:fig:f(H)}.

These calculations demonstrates that \emph{only the axial model with easy-axis anisotropy is in {qualitative} agreement with the experiment}. It predicts rather well defined softening of the resonance modes at spin-reorientation transition and it predicts mode with non-Larmor slope to originate from the smaller gap. This means that the larger gap $\Delta_2$ is of the exchange origin and corresponds to some oscillations strongly distorting spin helix.

Spin-reorientation occurs in this model for the field applied perpendicular to the symmetry axis $Z$: at field $H_c=\sqrt{\frac{\abs{A}}{\gamma^2 (I-I_n)}}$ plane of the spin structure suddenly rotates from the $(XY)$-plane  to the plane orthogonal to the magnetic field. Note, that contrary to the case of conventional easy-axis two-sublattices antiferromagnet spin-reorientation transition observed ah $\vect{H}\perp Z$, i.e. in the powder particles with high statistical weight, which facilitates observation of this transition both in static and ESR experiments. Also note, that for non-collinear antiferromagnet $\Delta_1 \neq \gamma H_c$. Critical field calculated for nabokoite for the model parameters described above is about 3.6~kOe, which exceeds observed value by approx. 50\%. This difference is most likely due to distortions of spin helix, since frequency of the exchange mode also changes significantly over this field range.

\end{document}